\documentclass[twoside,twocolumn,9pt]{article}
\usepackage{extsizes}
\usepackage[super,sort&compress,comma]{natbib} 
\usepackage[version=3]{mhchem}
\usepackage[left=1.5cm, right=1.5cm, top=1.785cm, bottom=2.0cm]{geometry}
\usepackage{balance}
\usepackage{times,mathptmx}
\usepackage{sectsty}
\usepackage{graphicx} 
\usepackage{lastpage}
\usepackage[format=plain,justification=justified,singlelinecheck=false,font={stretch=1.125,small,sf},labelfont=bf,labelsep=space]{caption}
\usepackage{float}
\usepackage{fancyhdr}
\usepackage{fnpos}
\usepackage[english]{babel}
\addto{\captionsenglish}{%
  
}
\usepackage{array}
\usepackage{droidsans}
\usepackage{charter}
\usepackage[T1]{fontenc}
\usepackage[usenames,dvipsnames]{xcolor}
\usepackage{setspace}
\usepackage[compact]{titlesec}

\usepackage{epstopdf}

\definecolor{cream}{RGB}{222,217,201}

\begin{document}

\clearpage

This document is the final version of the accepted manuscript in Nanoscale, doi: 10.1039/D0NR00185F. The quality of all figures is significantly reduced.

\pagestyle{fancy}
\thispagestyle{plain}

\makeFNbottom
\makeatletter
\renewcommand\LARGE{\@setfontsize\LARGE{15pt}{17}}
\renewcommand\Large{\@setfontsize\Large{12pt}{14}}
\renewcommand\large{\@setfontsize\large{10pt}{12}}
\renewcommand\footnotesize{\@setfontsize\footnotesize{7pt}{10}}
\makeatother

\renewcommand{\thefootnote}{\fnsymbol{footnote}}
\renewcommand\footnoterule{\vspace*{1pt}%
\color{cream}\hrule width 3.5in height 0.4pt \color{black}\vspace*{5pt}} 
\setcounter{secnumdepth}{5}

\makeatletter 
\renewcommand\@biblabel[1]{#1}            
\renewcommand\@makefntext[1]%
{\noindent\makebox[0pt][r]{\@thefnmark\,}#1}
\makeatother 
\renewcommand{\figurename}{\small{Fig.}~}
\sectionfont{\sffamily\Large}
\subsectionfont{\normalsize}
\subsubsectionfont{\bf}
\setstretch{1.125} 
\setlength{\skip\footins}{0.8cm}
\setlength{\footnotesep}{0.25cm}
\setlength{\jot}{10pt}
\titlespacing*{\section}{0pt}{4pt}{4pt}
\titlespacing*{\subsection}{0pt}{15pt}{1pt}

\fancyfoot{}
\fancyfoot[LO,RE]{\vspace{-7.1pt}\includegraphics[height=9pt]{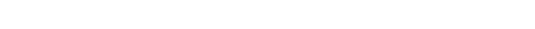}}
\fancyfoot[CO]{\vspace{-7.1pt}\hspace{13.2cm}\includegraphics{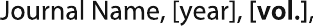}}
\fancyfoot[CE]{\vspace{-7.2pt}\hspace{-14.2cm}\includegraphics{head_foot/RF}}
\fancyfoot[RO]{\footnotesize{\sffamily{1--\pageref{LastPage} ~\textbar  \hspace{2pt}\thepage}}}
\fancyfoot[LE]{\footnotesize{\sffamily{\thepage~\textbar\hspace{3.45cm} 1--\pageref{LastPage}}}}
\fancyhead{}
\renewcommand{\headrulewidth}{0pt} 
\renewcommand{\footrulewidth}{0pt}
\setlength{\arrayrulewidth}{1pt}
\setlength{\columnsep}{6.5mm}
\setlength\bibsep{1pt}

\makeatletter 
\newlength{\figrulesep} 
\setlength{\figrulesep}{0.5\textfloatsep} 

\newcommand{\topfigrule}{\vspace*{-1pt}%
\noindent{\color{cream}\rule[-\figrulesep]{\columnwidth}{1.5pt}} }

\newcommand{\botfigrule}{\vspace*{-2pt}%
\noindent{\color{cream}\rule[\figrulesep]{\columnwidth}{1.5pt}} }

\newcommand{\dblfigrule}{\vspace*{-1pt}%
\noindent{\color{cream}\rule[-\figrulesep]{\textwidth}{1.5pt}} }

\makeatother

\twocolumn[
  \begin{@twocolumnfalse}
\vspace{3cm}
\sffamily
\begin{tabular}{m{4.5cm} p{13.5cm} }

\includegraphics{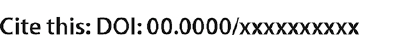} & \noindent\LARGE{\textbf{Epitaxial graphene/Ge interfaces: a minireview}} \\
\vspace{0.3cm} & \vspace{0.3cm} \\

 & \noindent\large{Yuriy Dedkov$^{\ast}$\textit{$^{a,b}$} and Elena Voloshina$^{\ast}$\textit{$^{a,b}$}} \\
 
\includegraphics{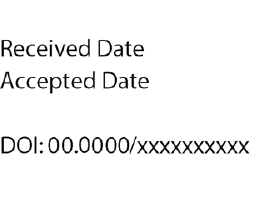} & \noindent\normalsize{The recent discovery of the ability to perform direct epitaxial growth of graphene layers on semiconductor Ge surfaces led to the huge interest to this topic. One of the reasons for this interest is the chance to overcome several present-day drawbacks on the way of the graphene integration in the modern semiconductor technology. The other one is connected with the fundamental studies of the new graphene-semiconductor interfaces, that might help with the deeper understanding of mechanisms, which governs graphene growth on different substrates as well as shed light on the interaction of graphene with these substrates, which range is now spread from metals to insulators. The present minireview gives a timely overview of the state-of-the-art in the field of studies of the graphene-Ge epitaxial interfaces and draw some perspective directions in this research area.} \\

\end{tabular}

 \end{@twocolumnfalse} \vspace{0.6cm}

]

\renewcommand*\rmdefault{bch}\normalfont\upshape
\rmfamily
\section*{}
\vspace{-1cm}


\footnotetext{\textit{$^{a}$~Department of Physics, Shanghai University, 200444 Shanghai, P. R. China.; E-mail: dedkov@shu.edu.cn, voloshina@shu.edu.cn}}
\footnotetext{\textit{$^{b}$~Institute of Physical and Organic Chemistry, Southern Federal University, 344090 Rostov on Don, Russia}}




\section*{Introduction}

Graphene (gr), a pure two-dimensional (2D) material formed by carbon atoms, already in 1947 was predicted to have extraordinary electronic properties~\cite{Wallace:1947dy}. For many years this material was considered as an interesting playground to study different theoretical phenomena~\cite{Semenoff:1984dq,Schakel:1991aa} assuming that it cannot exist in the free-standing form due to its 2D nature~\cite{Hohenberg:1967aa,Mermin:1968aa}. After graphene isolation in the free-standing form in 2004~\cite{Novoselov:2004a,Novoselov:2005es,Zhang:2005gp} and experiments on its intriguing transport, mechanical, optical, and other properties~\cite{Geim:2007hy,Geim:2009} there was a boom in the graphene properties research and the subsequent studies further evolve in the huge research area of 2D materials~\cite{Geim:2014hf,Duan:2015fr,Manzeli:2017ib}.

Besides that, in surface science, graphene (or monolayer of graphite in a former time) is known since the middle of 60s~\cite{Hagstrom:1965vh,May:1969uj,Land:1992,Tontegode:1991ts,Dedkov:2001,Batzill:2012,Dedkov:2015kp}. Initially, these mono- and multilayer graphitic layers on metal surfaces were considered as parasitic poison coatings, which block the catalytic activity of the metals. Later, it was found that graphene layer on metal itself is a very interesting object in surface and materials science as well as in graphene research. For example, graphene layers were predicted to act as spin filters in the FM/$n$-gr/FM sandwiches~\cite{Karpan:2007,Karpan:2008} (FM = ferromagnet, Ni or Co), graphene layers can be used as a protective inert cover for metals and semiconductors~\cite{Kubler:2005fm,Dedkov:2008d,Dedkov:2008e,Sutter:2010bx,Weatherup:2015cx}, graphene-based moir\'e structures on close-packed surfaces of $4d$ and $5d$ metals are considered as substrates for the growth of the ordered arrays of molecules or metallic clusters~\cite{NDiaye:2009a,Gerber:2013fa,Paschke:2019ei} with the perspective to use these systems in nano-catalysis or for data storage, gr-metal interfaces are considered as a perspective systems for the confined catalysis~\cite{Deng:2016ch,Fu:2016gt}, etc.

\begin{figure*}[t]
\centering
\includegraphics[width=\textwidth]{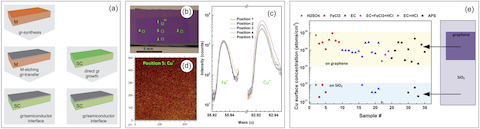}
\caption{(a) Schemes for the graphene fabrication on semiconductors: (left) using metal (M) substrate for graphene synthesis and then etching and transfer and (right) direct graphene growth on semiconductor (SC) using CVD or MBE methods. (b) Optical microscope image of a graphene layer transferred onto a $300$\,nm-SiO$_2$/Si substrate. (c) ToF-SIMS mass spectra in $^{56}\mathrm{Fe}^+$ and $^{63}\mathrm{Cu}^+$ regions acquired at different points in (b) across the sample. (d) $500\times500\mu \mathrm{m}^2$ ToF-SIMS map of Cu$^+$ in the center of the sample. (e) Comparison of surface concentration of Cu for different transfer methods. Measurements on graphene were performed in the center of the flake. Control measurements on the SiO$_2$ substrate were performed $0.5-1$\,mm away from the edge of the graphene flake. Images in (b-e) are reproduced from Ref.~\citenum{Lupina:2015je} with permission.}
\label{grM_grS_scheme}
\end{figure*}

From the technological point of view, the graphene synthesis on metals was recently considered as a most perspective one~\cite{Li:2009,Kim:2009a} and it was shown that huge mono- and bi-layer graphene sheets with size up to 30-inches can be synthesised on the polycrystalline Cu foil and then transferred on the polymer or semiconducting support for further processing (Fig.~\ref{grM_grS_scheme}(a))~\cite{Bae:2010}. Later this method was successfully applied for the fabrication of the first graphene-based touch screens for mobile phones~\cite{Ryu:2014fo}.

However, the technological adaptation of such synthesis of graphene on metals followed by the transfer on the desired support has several drawbacks. First one is the low scalability of this method, which does not allow controllably produce small elements from graphene of desired shape and edge configuration. The second problem is connected with the use of different chemical reagents and polymers used in the transfer process. Also the used support always contains unavoidable contaminants. Together, these factors lead to the fabrication of the contaminated graphene-support interface, that can drastically modify the doping level of graphene and, hence, its transport properties. The third main drawback after the metal-based graphene synthesis is connected with the residual metal contamination (Cu, in most cases, as a widely used substrate for the mass production) of a graphene layer. Recent experiments demonstrate that trace amounts of metals ($\sim10^{13}-10^{14}\,\mathrm{atoms}/\mathrm{cm}^2$) are found on graphene transferred to the target SiO$_2$/Si wafer~\cite{Ambrosi:2014gl,Lupina:2015je} (Fig.~\ref{grM_grS_scheme}(b-e)). It was found that even such small amounts may be relevant during front-end-of-line integration approaches and can lead to the contamination of Si-based devices and cross-contamination of fabrication tools. Therefore the search for the new graphene synthesis methods directly on semiconductor surfaces was very active in the last decade. In such a way synthesised graphene layers, either after their transfer on the desired semiconductor support or directly grown on semiconductors, can be used in different attractive applications like, for example, photodetectors~\cite{XiaohongAn:2013kq,Liu:2014daa}, chemical sensors~\cite{HyeYoungKim:2013bj,Singh:2013dh}, mixers~\cite{Gu:2012jo}, optical modulators~\cite{Liu:2011ex}, solar cells~\cite{Li:2010iw,An:2013gj}, etc. In all these devices, the so-called Schottky-barrier at the graphene-semiconductor interface is formed~\cite{DiBartolomeo:2016ii}, attracting much attention in fundamental and application-like studies.

Initial experiments on the direct graphene synthesis on Si and SiO$_2$/Si surfaces using chemical vapour deposition (CVD) or using the solid source in molecular beam epitaxy (MBE) demonstrated that in most cases such deposition of carbon leads to the transformation of the surface layers of the underlying Si substrate into SiC layer~\cite{Hackley:2009bf,ThanhTrung:2013io,ThanhTrung:2014hd,Maeda:2011bt}. Only recently, it was demonstrated, that after careful adjustment of the synthesis parameters, graphene layers can be grown via CVD directly on the Si substrate and the formation of the Si-C bond in this case was suppressed~\cite{Kim:2011fs,Tai:2018fx}. 

The further progress in the search of the appropriate methods and semiconductor substrates led to the discovery of the direct graphene CVD synthesis on Ge surfaces in 2013~\cite{Wang:2013fq}, followed by the successful implementation of this approach for the wafer-scale graphene preparation~\cite{Lee:2014dv}.  These findings stimulated a rapid growth of many experimental and theoretical works in this area. Here we present a timely review of the recent progress in the studies of the graphene growth on Ge semiconductor substrates, crystallographic structures and electronic properties of the formed graphene-Ge interfaces with different substrate orientations. In the end we present some ideas and perspectives on how properties of such interfaces might be tailored for the further use in mico(nano)electronics and spintronics.

\section*{Graphene structure on Ge surfaces}

\begin{figure*}[t]
\centering
\includegraphics[width=\textwidth]{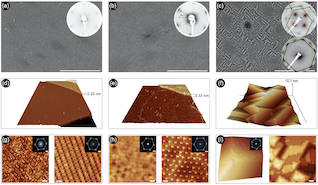}
\caption{Upper row: SEM images of the morphology of continuous graphene layer and the underlying substrate on: (a) Ge(110), (b) Ge(111), (c) Ge(001). Scale bars are $1\,\mu\mathrm{m}$. Insets in (a-c) show LEED patterns of single-crystal graphene grown on respective Ge surfaces. Upper/lower insets in (c) show LEED patterns directly after graphene growth on Ge(001) and after additional annealing at $850^\circ$\,C under Ar-atmosphere of $800$\,mbar. Middle row: Large scale STM images of (d) gr/Ge(110) (size: $500\,\mathrm{nm} \times 310\,\mathrm{nm}$, $U_T = 2\,\mathrm{V}$, $I_T = 100\,\mathrm{pA}$), (e) gr/Ge(111)  (size: $200\,\mathrm{nm} \times 163\,\mathrm{nm}$, $U_T = -1\,\mathrm{V}$, $I_T = 100\,\mathrm{pA}$), (f) gr/Ge(001) (size: $200\,\mathrm{nm} \times 142\,\mathrm{nm}$, $U_T = 2\,\mathrm{V}$, $I_T = 200\,\mathrm{pA}$). Lower row: corresponding atomically resolved STM images acquired before (left) and after (right) UHV annealing at $700^\circ$\,C for (g) gr/Ge(110), (h) gr/Ge(111), and (i) gr/Ge(001). Insets show corresponding FFT images. Scale bars are $2$\,nm and $4$\,nm for (g-h) and (i), respectively. Images are reproduced from Refs.~\citenum{Lee:2014dv,Kiraly:2015kaa,Dabrowski:2017gr,Sitek:2020hw} with permission.}
\label{grGe_SEM_LEED_STM}
\end{figure*}

First experiments on the direct growth of graphene on Ge substrates using CVD (mixture of CH$_4$, H$_2$, and Ar) and MBE (carbon atomic source) methods were published in 2013-2014~\cite{Wang:2013fq,Lee:2014dv,Lippert:2014fc}. It was shown that all main surfaces of Ge -- (001), (110), and (111) -- can be used for the graphene growth (Fig.~\ref{grGe_SEM_LEED_STM}); however, the graphene and interfaces quality as well as the system morphology are different in all cases. 
The success of these experiments is based on the catalytic activity of Ge surfaces and that under equilibrium conditions the parts of the Ge-C alloy do not intermix with each other~\cite{Wang:2013fq}. This is similar to the conditions for the graphene growth on Cu surfaces. Therefore, taking into account the extremely small solubility of carbon in bulk Ge ($<0.1$\%), one can conclude that during CVD process the graphene growth is self-limiting and surface-mediated - similar to the Cu-assisted growth and contrary to the Ni-catalysed growth, where carbon has a high solubility and several competitive processes take place. It was shown that for the graphene growth on Ge the fast or slow cooling rates do not play a crucial role, but the temperature during graphene synthesis by means of CVD or MBE is a key factor, which determines the alignment of graphene fragments on Ge surfaces and later the agglomeration to the complete layer. 

Figure~\ref{grGe_SEM_LEED_STM} shows $\mu\mathrm{m}$-scale scanning electron microscopy (SEM) images of (a) gr/Ge(110), (b) gr/Ge(111), and (c) gr/Ge(001)~\cite{Sitek:2020hw}. These data are accompanied by the respective low-energy electron diffraction (LEED) images shown as insets (in case of gr/Ge(001), upper one shows LEED directly after graphene growth and lower one after additional post-annealing). All experimental works indicate that for all Ge surfaces, temperature, which is used during graphene synthesis by means of CVD or MBE, is a crucial factor~\cite{Lee:2014dv,Rogge:2015hl,DiGaspare:2018ha,Tesch:2018hm,Persichetti:2019hn,Persichetti:2020il,Sitek:2020hw}. As was found the quality of graphene improves dramatically as the temperature used during synthesis approaches $930^\circ$\,C and the temperature decrease of only $10^\circ$\,C already leads to a wrinkled and defective graphene layers as demonstrated using scanning tunnelling microscopy (STM) and Raman spectroscopy~\cite{Persichetti:2019hn,Persichetti:2020il,Sitek:2020hw}. This abrupt change in the graphene quality and system morphology is connected with the formation of the quasi-liquid surface layer of Ge at temperatures several degrees below melting point. This leads to the higher mobility and sublimation rate on the Ge surface that in its turn promotes the high diffusion of carbon species and the desorption of the defective fragments from the sample surface. 
However, in order to better understand the conditions for the high-quality graphene growth, further experimental and, especially, theoretical studies on the growth kinetics of graphene on Ge surfaces are necessary.

The morphology analysis performed with SEM for different gr/Ge interfaces reveals that the surface is flat for the case of Ge(110) and Ge(111) (Fig.~\ref{grGe_SEM_LEED_STM}(a,b)). This fact is confirmed by formation of hexagonal LEED patterns (shown as corresponding insets) as well as by STM data (Fig.~\ref{grGe_SEM_LEED_STM}(d,e)). For gr/Ge(110) the single-domain graphene growth is observed at high temperatures as found by LEED and transmission electron microscopy (TEM)~\cite{Lee:2014dv,Tesch:2018hm}, whereas for gr/Ge(111) the polycrystalline graphene layer is formed with a weak domain orientation preference. In case of the gr/Ge(001) interface, the strong faceting of the Ge surface under graphene is observed with the preferential formation of the $\{107\}$ facets~\cite{Kiraly:2015kaa,McElhinny:2016gw,Pasternak:2016ec,Lukosius:2016ce,Dabrowski:2017gr,DiGaspare:2018ha,Persichetti:2019hn,Sitek:2020hw}, as it is revealed in SEM and STM experiments (Fig.~\ref{grGe_SEM_LEED_STM}(c,f)). LEED patterns for this system acquired after synthesis demonstrate 3 pairs of hexagons due to facets on the Ge(001) surface and in every pair the spots are rotated by $30^\circ$ with respect to each other due to the existence of two planar graphene orientations (upper inset of (c)). Such strong difference in the graphene morphology is assigned to the fact that on Ge(110) the growth of graphene is anisotropic with the graphene islands uniaxially aligned along the Ge\,$\langle\overline{1}10\rangle$ direction (``zig-zag'' and ``arm-chair'' graphene edges are parallel to Ge\,$\langle001\rangle$ and Ge\,$\langle\overline{1}10\rangle$, respectively), whereas for Ge(111) and Ge(001) the graphene seeds shapes are isotropic, that leads to the polycrystalline growth~\cite{Lee:2014dv,Dai:2016jm,Sitek:2020hw}. Density functional theory (DFT) calculations support these experimental findings giving lowest formation energy for graphene fragments aligned with ``arm-chair'' edges along the Ge steps parallel to the $\langle\overline{1}10\rangle$ direction~\cite{Dai:2016jm}. The observed graphene islands alignment and the growth anisotropy for the Ge(110) surface leads to the appearance of the barrier for the islands rotation during growth that finally leads to the single-domain graphene layer on this surface.

Graphene growth on Ge surfaces leads to the respective changes of the structure of the underlying surfaces (Fig.~\ref{grGe_SEM_LEED_STM}). Available experimental data give clear information about atomic arrangement at the gr/Ge(110) interfaces with less know structures for other two gr/Ge systems. 

\begin{figure*}
\centering
\includegraphics[width=\textwidth]{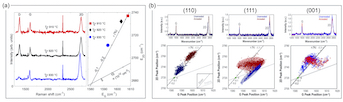}
\caption{(a) Raman spectra of graphene samples grown on Ge(110) at different temperatures. (b) Upper row: Raman spectra of graphene on Ge(110), Ge(111), and Ge(001) before (blue) and after (red) UHV annealing at $700^\circ$\,C. Bottom row: Corresponding plot of Raman 2D peak position versus G peak position for spatially resolved Raman spectra of graphene on Ge surfaces before (blue) and after (red) vacuum annealing. Shaded circles indicate 95\% confidence intervals. Shifts along the dark grey lines are associated with strain in the graphene lattice, while shifts along the light grey line are associated with variations in carrier doping. Images in (a) and (b) are reproduced from Ref.~\citenum{Persichetti:2020il} and Ref.~\citenum{Kiraly:2015kaa}, respectively, with permission.}
\label{grGe_RamanSpectr}
\end{figure*}

The remarkable example of the modification of the underlying substrate after graphene growth is the gr-Ge(001) interface, where, as discussed, graphene growth leads to the strong surface faceting. However, it was demonstrated by LEED and STM that ultra-high vacuum (UHV) annealing at $700^\circ$\,C, or annealing at $850^\circ$\,C under Ar-atmosphere of $800$\,mbar, or slow sample cooling after growth leads to the flattening of the gr-Ge(001) interface and some areas demonstrate $(2\times1)$ reconstruction, representative for freshly UHV-prepared Ge(001) (see lower inset of Fig.~\ref{grGe_SEM_LEED_STM}(c) and right image in Fig.~\ref{grGe_SEM_LEED_STM}(i))~\cite{Kiraly:2015kaa,Dabrowski:2017gr,McElhinny:2016gw}. Therefore, we can conclude that fast sample cooling after graphene growth simply drives the quenching of facets on the Ge surface, which are formed under graphene and which can be either avoided by slow cooling rate or using additional postannealing.

The not so dramatic, but also notable changes appear at the Ge(110) and Ge(111) surfaces after graphene growth and the following postannealing procedures. As was found, the CVD graphene growth on Ge(110) leads to the disordered structure of the Ge surface quenched by graphene (Fig.~\ref{grGe_SEM_LEED_STM}(g), left)~\cite{Kiraly:2015kaa,Dai:2016jm,Kiraly:2018db}. This can be compared to the coexisting $c(8\times10)$ and $(16\times2)$ reconstructions characteristic for pristine Ge(110). Annealing of the as-grown gr/Ge(110) at $700^\circ$\,C in UHV drives the formation of the $(6\times2)$-Ge(110) reconstruction, which is stabilized by the graphene layer and which was not previously reported for this Ge surface (Fig.~\ref{grGe_SEM_LEED_STM}(g), right)~\cite{Kiraly:2015kaa,Kiraly:2018db,Campbell:2018iia,Zhou:2018ji}. As shown by STM and surface x-ray diffraction (SXRD), the interfaced Ge layer reorganizes in clusters, which are ordered along the $\langle\overline{1}12\rangle$ direction of bulk Ge. It is interesting to note, that the similar $(6\times2)$-Ge(110) surface under graphene is formed after direct MBE graphene growth on the same surface~\cite{Rogge:2015hl,Tesch:2017gm,Tesch:2018hm}. 

For the gr/Ge(111) interface prepared by CVD the topmost Ge layer does not show any specific reconstruction (Fig.~\ref{grGe_SEM_LEED_STM}(h), left), which, however, transformed to the 6-fold ordered structure after UHV annealing at $700^\circ$\,C (Fig.~\ref{grGe_SEM_LEED_STM}(h), right) and the area for this structure is increased from less than $20\%$ to $100\%$ upon thermal treatment~\cite{Kiraly:2015kaa}. The formed reconstruction of Ge(111) under graphene deviates from the representative $c(2\times8)$ structure for pristine Ge(111). For the MBE grown gr/Ge(111) interface the recovery of the typical $c(2\times8)$-Ge(111) reconstruction under graphene was detected upon sample cooling, pointing the importance of the used graphene synthesis methods~\cite{Rogge:2015hl}. It was also found that all reconstructed gr/Ge interfaces survived the ambient exposure that gives a strong support for the conclusion that graphene in these systems modifies the energetic landscape of the interfaced Ge surfaces.

\section*{Electronic properties of graphene on Ge surfaces}

Graphene growth conditions, like the ratio of the CH$_4$ flow with respect to the ones for H$_2$ and Ar as well as the used growth temperatures, have a dramatic implication on the crystallographic quality of a graphene layer on Ge and, respectively, on its electronic properties. Figure~\ref{grGe_RamanSpectr}(a)~\cite{Persichetti:2020il} shows Raman spectra measured for a series of gr/Ge(110) prepared at different substrate temperatures (left panel) and the respective peak energy position diagram (right panel), which is used for the analysis of the strain and doping level in graphene. The summary of the space-resolved data for graphene layers prepared on different Ge substrates and measured before and after UHV annealing at at $700^\circ$\,C is shown in panel (b)~\cite{Kiraly:2015kaa}, respectively. All spectra reveal the main feature of graphene, namely 2D ($\sim2700\,\mathrm{cm}^{-1}$) and G ($\sim1600\,\mathrm{cm}^{-1}$) bands. Also the D peak ($\sim1350\,\mathrm{cm}^{-1}$), which originates from the intervalley resonant scattering processes induced by defects, can be also recognised in the spectra. [Sharp features at $\sim1550\,\mathrm{cm}^{-1}$ and $\sim2330\,\mathrm{cm}^{-1}$ are attributed to ambient oxygen and nitrogen, respectively.]

In all studied cases for the preparation of gr/Ge(110)~\cite{Wang:2013fq,Zhao:2019eh,Persichetti:2020il} and gr/Ge(001)~\cite{Wang:2013fq,Pasternak:2016ec,Lukosius:2016ce,Persichetti:2019hn}, the increase of the synthesis temperature between $910^\circ$\,C and $930^\circ$\,C leads to the abrupt changes in the quality of graphene as indicated by the significant reduction of the intensity of the D peak as well as by the increase of the ratio of intensities of the 2D and G peaks (Fig.~\ref{grGe_RamanSpectr}(a)). These observations can be clearly assigned to the decrease of the number of defects in the formed graphene layer. The analysis of the 2D vs G band energies~\cite{Lee:2012gy} for the space-integrated data shows that the increase of the synthesis temperature leads to the decrease of the compressive strain in graphene from $\approx0.5\%$ to $\approx0.3\%$ with the negligible doping for gr/Ge(110)~\cite{Persichetti:2020il}; the opposite increase of the strain is observed for gr/Ge(001) with the electron doping density in graphene of $\approx10^{13}\mathrm{cm}^{-2}$~\cite{Pasternak:2016ec,Lukosius:2016ce,Dabrowski:2016im,Persichetti:2019hn}. 

\begin{figure*}
\centering
\includegraphics[width=\textwidth]{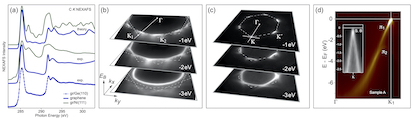}
\caption{(a) Experimental C $K$-edge NEXAFS spectra of gr/Ge(110), graphene, and gr/Ni(111). The respective theoretical spectra for the last two samples are shown at the top of the panel. Constant energy cuts (CECs) of the ARPES intensity for (b) two-domain and (c) single-domain gr/Ge(110). Dashed-line hexagons mark the respective hexagonal Brillouin zones of graphene for both samples. (d) ARPES intensity map presented along the $\Gamma-\mathrm{K}_1$ direction (marked in (b)) of the graphene-derived Brillouin zone of one of the graphene domains. The inset of (d) shows the photoemission intensity cut along the direction perpendicular to $\Gamma-\mathrm{K}$ (marked in (c)). All ARPES data were collected at $T = 100$\,K with a photon energy of $h\nu = 100$\,eV. Images are reproduced from Ref.~\citenum{Tesch:2018hm} with permission.}
\label{grGe_NEXAFS_ARPES}
\end{figure*}

The interesting observations were made in the Raman spectroscopy experiments for the gr/Ge samples after UHV annealing at $700^\circ$\,C (Fig.~\ref{grGe_RamanSpectr}(b))~\cite{Kiraly:2015kaa}. In case of gr/Ge(110) and gr/Ge(111) the increase of the compressive strain was observed, that was connected with the formation of the reconstructed $(6\times2)$-Ge(110) under graphene and with the increase of the chemical interaction between graphene and Ge(111), respectively. The later observation is also supported by the scanning tunnelling spectroscopy (STS) data for the gr/Ge(111) interface indicating the discussed changes. For the gr/Ge(001), the experimental data shows significant point-to-point energy variation of the peak, however, the compressive strain of $\approx0.5\%$ was measured which does not change upon vacuum annealing. These data from Ref.~\citenum{Kiraly:2015kaa}, however, are in contradiction with the recently published data for this interface~\cite{Sitek:2020hw}, which demonstrate no significant dispersion in the peak position with a compressive strain of $\approx0.1\%$. These discrepancies are attributed to the use of the doped Ge substrates in the former studies~\cite{Kiraly:2015kaa}, indicating the importance of further works of the substrate doping on the graphene electronic properties.

The number of the available electronic structure studies by electron spectroscopy methods and DFT is very limited and here we discuss almost all present-day results. The interaction of graphene with Ge substrates can be characterised, on the first step, with x-ray photoelectron spectroscopy (XPS) and near-edge x-ray absorption fine structure spectroscopy (NEXAFS). The presence of the single C\,$1s$ component in XPS spectra confirms the homogeneity of the $sp^2$ phase in the formed graphene layers on different Ge surfaces and the XPS results, measured for the UHV annealed gr/Ge samples, indicate the weak $n$-doping of graphene on Ge(110) and Ge(001) with a position of the C\,1s peak in the range of $284.4-285.0$\,eV~\cite{Lee:2014dv,Pasternak:2016ec,Scaparro:2016jc,Tesch:2017gm,Tesch:2018hm,DiGaspare:2018ha,Zhou:2018ji,Persichetti:2019hn} as compared to $284.23$\,eV for the neutral graphene or graphite~\cite{Preobrajenski:2008,Schroder:2016eb}. NEXAFS spectra measured at the C\,$K$ absorption edge for gr/Ge(110) (Fig.~\ref{grGe_NEXAFS_ARPES})~\cite{Tesch:2017gm,Tesch:2018hm} are in very good agreement with the experimental and theoretically calculated ones for free-standing graphene (graphite)~\cite{Preobrajenski:2008,Voloshina:2013cw}, indicating very weak interaction between a graphene layer and Ge substrate. [NEXAFS intensity peaking at $285.3$\,eV is assigned to the excitation of the $1s$ electron onto the $\pi^*$ unoccupied states of graphene and intensity in the range of $291-295$\,eV to the $1s\rightarrow\sigma^*$ transition.] Here, one can compare these results with the NEXAFS spectra of the \textit{strongly} interacting gr/Ni(111) interface (shown in the figure), where substantial modifications of the NEXAFS spectrum compared to the one for free-standing graphene is observed~\cite{Weser:2010,Dedkov:2010jh,Rusz:2010}.

First published angle-resolved photoelectron spectroscopy (ARPES) data were obtained using nano- and $\mu$-ARPES on graphene flakes prepared on Ge(001) by means of CVD and MBE~\cite{Dabrowski:2016im}. In consistence with the previous LEED and SEM data, it was found that 12 Dirac cones are observed in the photoemission full data sets corresponding to two-domains growth mode for the gr/Ge(001) interface. Here the electronic structure of graphene was found almost intact with a rigid shift of bands indicating $p$-doping with a position of the Dirac point $E_D-E_F=+0.185$\,eV. Further studies of the gr/Ge(001) interface prepared by CVD method~\cite{Dabrowski:2017gr} found $n$-doping of graphene with the Dirac point positions $E_D-E_F=-0.05$\,eV and $E_D-E_F=-0.1$\,eV for as-prepared and for Ar-atmosphere-annealed at $850^\circ$\,C samples, respectively (corresponding LEED images are presented as insets in Fig.~\ref{grGe_SEM_LEED_STM}(c)). These new results point out the absence of H at the gr/Ge interface as was proposed in earlier works~\cite{Lee:2014dv}. Also, one can conclude that the previously observed in Ref.~\citenum{Dabrowski:2016im} the $p$-doping of graphene might be due to the residual ambient contaminations or partly intercalated oxygen or other species that was avoided by the additional UHV annealing of the studied gr/Ge(001) samples in Ref.~\citenum{Dabrowski:2017gr}.

The electronic structure of gr/Ge(110) prepared by MBE method and annealed in UHV conditions at $800^\circ$\,C was systematically studied by means of ARPES in Ref.~\citenum{Tesch:2018hm} and these results are summarized in Fig.~\ref{grGe_NEXAFS_ARPES}(b-d). According to this work, two gr/Ge(110) samples -- two-domains and single-domain -- prepared at slightly different temperatures, demonstrate $n$-doping of a graphene layer with the Dirac point position $E_D-E_F=-0.21$\,eV. These results are supported by the local STS $dI/dV$-mapping data, which give $E_D-E_F\approx-0.25$\,eV. The same gr/Ge(110) interface, but prepared by means of CVD and annealed in UHV conditions at $900^\circ$\,C, shows almost neutral state for graphene~\cite{Ahn:2018fja}. Further annealing of the same sample at $925^\circ$\,C leads to the shift of the Dirac point to $E_D-E_F=-0.1$\,eV, that was assigned to the deintercalation of H from the gr/Ge interface. This was conformed by further H-intercalation in gr/Ge(110) and restoring of the neutral state of graphene~\cite{Ahn:2018fja}. It is interesting to note, that low-temperature ARPES ($100$\,K) and STS $dI/dV$-mapping ($10$\,K) data give a relatively high value for the Fermi velocity in graphene -- $v_F=(1.38\pm0.15)\times10^6\,\mathrm{m\,s}^{-1}$  and $v_F=(1.82\pm0.21)\times10^6\,\mathrm{m\,s}^{-1}$, respectively, compared to $v_F=0.85\times10^6\,\mathrm{m\,s}^{-1}$ obtained within the local-density approximation (LDA) in the DFT for the case of fully screened electron-electron interaction in graphene. As concluded in Ref.~\citenum{Tesch:2018hm} it is due to the semiconducting Ge substrate used for the graphene growth, which has non-infinite dielectric constant, that in this case leads to the Fermi velocity renormalization and these results are in agreement with the recent ARPES data for graphene on other semiconducting or insulating substrates~\cite{Hwang:2012he,Ryu:2017dx}.

\begin{figure}
\centering
\includegraphics[width=\columnwidth]{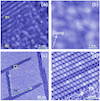}
\caption{(a) STM image of R1 and R2 phases of gr/Ge(110). (b) STM zoom-in scan of phase R2 showing the honeycomb lattice of graphene ($U_T = -0.2\,\mathrm{V}$, $I_T = 400\,\mathrm{pA}$). (c) STM image of gr/H/Ge(110) after prolonged UHV annealing at $300^\circ$\,C ($U_T = 2.5\,\mathrm{V}$, $I_T = 1\,\mathrm{nA}$). (d) STM zoom-in scan marked by yellow rectangle in (c) ($U_T = -2.5\,\mathrm{V}$, $I_T = 1\,\mathrm{nA}$). Images are reproduced from Ref.~\citenum{Zhou:2018ji} with permission.}
\label{grGe_110Hint}
\end{figure}

DFT analysis of the discussed gr/Ge interfaces is quite difficult due to the unknown interface structures. Therefore, the initial attempts in the recent theoretical works were devoted to the simulation of the electronic band structure of these systems using either bulk-terminated or known surface reconstructions of Ge surfaces. For example, the binding energy of graphene to Ge surface is $E_B=-40$\,meV/C-atom and $E_B=-37$\,meV/C-atom for gr/Ge(111)$c(2\times8)$ and gr/Ge(110)$c(8\times10)$, respectively, where corresponding original Ge surface reconstructions are considered~\cite{Rogge:2015hl}. The gr-Ge-adatom distances in these results are $\approx3$\,\AA. For the bulk terminated Ge surface in gr/Ge(110) the value of $E_B=-42$\,meV/C-atom with the gr-Ge-plane mean distance of $3.54$\,\AA~\cite{Tesch:2018hm}. The small difference between binding energies obtained for different models of the gr/Ge(110) interface is due to the weak C-Ge interaction and to the fact that no Ge-C alloy can be formed according to the phase diagram. According to the results presented in Ref.~\citenum{Tesch:2018hm}, graphene is weakly $p$-doped for the pure unreconstructed gr/Ge(110) interface with the Dirac point position of $E_D-E_F\approx0.2$\,eV. In order to reproduce the experimentally observed $n$-doping of graphene, the interfacial Sb atoms were used, which, as suggested might segregate at the gr-Ge interface during sample preparation. However, the experimental XPS data do not support this model as the detected amount of Sb atoms is much lower compared to the concentration used in the DFT calculations. As suggested in Ref.~\citenum{Ahn:2018fja}, the unsaturated dangling bonds of the Ge atoms at the gr-Ge interface might donate electrons to graphene inducing its $n$-doping observed in the experiment. Furthermore, more realistic models, which take into account the experimentally observed reconstructions at the gr-Ge interfaces, have to be considered for the correct description of the electronic properties of these interfaces. Also, such important factor as the doping of Ge substrate has to be considered as it might strongly influence the properties of the synthesised graphene layer.

\begin{figure*}
\centering
\includegraphics[width=\textwidth]{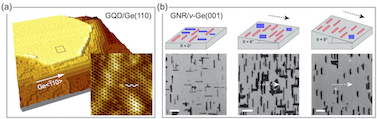}
\caption{(a) 3D representation of GQD/Ge(110) (size: $80\,\mathrm{nm} \times 80\,\mathrm{nm}$, $U_T = 2.4\,\mathrm{V}$, $I_T = 1\,\mathrm{nA}$). Inset shows a zoom-in STM scan marked by black rectangle in large scale scan (size: $5\,\mathrm{nm} \times 5\,\mathrm{nm}$, $U_T = 0.2\,\mathrm{V}$, $I_T = 2\,\mathrm{nA}$). Graphene ``arm-chair'' and Ge\,$\langle\overline{1}10\rangle$ are marked in the respective images. (b) Schematic diagrams (upper row) and SEM images (bottom row) of graphene crystals grown on Ge(001) with $0^\circ$, $6^\circ$, and $9^\circ$ miscut toward Ge\,$\langle110\rangle$. Red and blue crystals in are perpendicular and parallel, respectively, to the miscut. Dashed arrows point downhill. Scale bars in SEM images are $1\,\mu\mathrm{m}$. Images in (b) are reproduced from Ref.~\citenum{Jacobberger:2019gk} with permission.}
\label{grGe_GD_GNR}
\end{figure*}

\section*{Graphene hetero- and nano-structures on Ge surfaces}

The electronic structure of graphene on different substrates can be modified using different approaches, like, e.\,g., adsorption of different atoms on top of graphene, intercalation of different species between a graphene layer and substrate, or via synthesis of low-dimensional graphene quantum-dots (GQDs) or graphene nanoribbons (GNRs), where properties of new quantum objects are defined by confinement effects~\cite{Batzill:2012,Dedkov:2015kp}. It is known that the band gap of GNRs depends on their width and edge structure with its maximum values for GNRs width below $10$\,nm. Here semiconducting GNRs can reach high carriers mobility, current density, thermal and electrical conductivity. However, to be used in the semiconducting technology, such GNRs have to be highly aligned in one direction on the semiconducting support. For example, in case of field-effect transistors (FETs), the use of unaligned ribbons can increase the conduction way, decreasing on/off conductance ration of FET. Therefore, the alignment of semiconducting GNRs on large scale Ge surfaces could be a significant step for integration of graphene in the modern semiconductor technology and electronics.

First experiments on intercalation in gr/Ge interfaces were performed using hydrogen as an intercalant, motivated by the previous works on the gr/H/SiC(0001) systems, where the electronic decoupling of graphene from the SiC substrate was found with the simultaneous improvement of the transport properties of graphene~\cite{Riedl:2009el,Riedl:2010du,Sforzini:2015cp}. The hydrogen intercalation in gr/Ge(001) was performed during sample cooling after sample growth under $800$\,mbar H$_2$ pressure using starting temperatures of $600-900^\circ$\,C~\cite{Grzonka:2018aaa,Judek:2019ha}. Hydrogen intercalation leads to the flattening of the initially faceted gr/Ge(001) interface as confirmed by LEED and SEM measurements. At the same time this leads to the increase of the compressive strain in a graphene layer which initially mimics reconstructed Ge surface. The intercalation of hydrogen also increases the number of defects in graphene. All together these factors explain the observed degraded by factor of $2$ the charge mobility for the gr/H/Ge(001) samples, compared to pristine samples~\cite{Grzonka:2018aaa}. 

The intercalation of hydrogen in gr/Ge(110) was systematically studied by means of STM~\cite{Zhou:2018ji} and ARPES~\cite{Ahn:2018fja}. The initial motivation for these works was a question on the role of hydrogen during the CVD growth of graphene on Ge surfaces and presence of H-atoms at the gr/Ge interface after synthesis. It was found that after graphene synthesis, two distinct phases are present at the surface (Fig.~\ref{grGe_110Hint}(a,b))~\cite{Zhou:2018ji}: R1 - graphene on $(6\times2)$-Ge(110) (discussed before) and R2 - identified later as a graphene layer on H-terminated Ge(110), i.\,e. on the $(1\times1)$ surface. The R1 phase can be fully converted to R2 one (and vice versa) via intercalation (de-intercalation) of hydrogen. The intermediate state of the H-de-intercalation is shown in Fig.~\ref{grGe_110Hint}(c,d)~\cite{Zhou:2018ji}, where STM images of the initially fully H-intercalated gr/Ge(110) system are shown after prolonged UHV annealing at $300^\circ$\,C; such treatment of gr/H/Ge(110) leads to the desorption of H-atoms from interface and finally R2 phase is fully converted to R1 phase. ARPES experiments give the position of the Dirac point $E_D-E_F=-0.102$\,eV and $E_D-E_F=-0.038$\,eV for gr/Ge(110) (R1) and gr/H/Ge(110) (R2), respectively~\cite{Ahn:2018fja}.

The electrochemical oxidation of the gr/Ge(110) interface was performed in Ref.~\citenum{BraeuningerWeimer:2019kx} with the goal to stabilise a graphene-protected GeO$_2$. It was possible to prepare $350$\,nm-thick GeO$_2$ layer at the gr/Ge(110) interface via electrochemical intercalation in a $0.25$\,N solution of anhydrous sodium acetate in glacial acetic acid. This treatment led to the significant enhancement of the intensity of graphene-related 2D and G Raman peaks without increase of the intensity of the graphene D peak, meaning that number of defects is not increased after intercalation. As a result, the compressive strain of graphene observed in gr/Ge(110) is fully relaxed after intercalation and very weak $p$-doping of graphene was observed in ARPES measurements. This approach might open some perspectives on the graphene stabilisation and use of thin GeO$_2$ oxides in future applications. The similar results, but for gr/Ge(001), were obtained on the studies of the oxidation and ageing effects of this interface~\cite{Dabrowski:2019hj,Mendoza:2019de}.

GQDs are small graphene fragments (typically with the lateral size below $50$\,nm), where electronic transport is confined in all three spatial dimensions. GQDs can be fabricated by fragmentation or ``cutting'' of graphene sheets (top-down approach). Alternatively, large graphene-like molecules can be synthesised with well-defined molecular structure (bottom-up approach). Figure~\ref{grGe_GD_GNR}(a) shows single GQD grown on Ge(110) at $910^\circ$\,C using low-pressure CVD method (partial pressure of C$_2$H$_4$ is $1\times10^{-4}$\,mbar) under UHV conditions. The main prerequisites here for the growth of high-quality GQDs with well-ordered edges are the initial preparation of the atomically flat Ge surface under UHV conditions and the low graphene growth rate. Considering the presented example one can see that Ge(110) surface behaves as a ``template'' defining the orientation of GQDs and here well-ordered ``arm-chair'' graphene edges of the island are aligned along the Ge\,$\langle\overline{1}10\rangle$ directions, as was previously observed for the complete graphene layers.

In Ref.~\citenum{Jacobberger:2015de} and \citenum{Saraswat:2019ei} the direct CVD growth at $860^\circ<T<935^\circ$\,C of ``arm-chair'' GNRs on Ge(001) and $3\mu\mathrm{m}$-Ge(001)/Si(001) was demonstrated. In these works the GNRs are self-aligned $3^\circ$-off with respect to the Ge\,$\langle110\rangle$ directions with predominantly smooth ``arm-chair'' graphene edges. Via tuning the growth parameters (temperature as well as the CH$_4$ and H$_2$ partial pressures) it was possible to reach very slow growth rate of $<5\,\mathrm{nm\,h}^{-1}$ in the GNRs' width direction. In this case it was possible to have a tuneable width for GNRs of below $10$\,nm and aspect ratio more than $70$. Taking into account that Ge(001) surface always presents two equivalent domains rotated by $90^\circ$ with respect to each other, the two equivalent GNRs arrays were observed. To overcome this issue, the use of vicinal Ge surfaces was proposed~\cite{Jacobberger:2019gk,Li:2019bo}. Using $v$-Ge(001) surface with different miscut angles towards Ge\,$\langle110\rangle$ it was possible to reach almost complete anisotropic ``arm-chair'' GNRs growth (Fig.~\ref{grGe_GD_GNR}(b))~\cite{Jacobberger:2019gk}. For miscut angles $0^\circ$, $6^\circ$, and $9^\circ$ the alignment probability for GNRs perpendicular to miscut of $50\%$, $73\%$, and $90\%$ was achieved, respectively~\cite{Jacobberger:2019gk}. For $v$-Ge(001)-$12^\circ$ the alignment probability for GNRs reaches $100\%$~\cite{Li:2019bo}. Using the same approach for the slow growth rate the high-quality GNRs with width below $10$\,nm and high aspect ratio can be grown. At the same time, if the growth rate is increased, i.\,e. the ratio for the flow rates of H$_2$ and CH$_4$ decreases, then for vicinal Ge(001) surfaces the downhill crystal edges of GNRs become longer than the uphill edges. These observations are explained by the stronger binding between the downhill edge of GNRs and the Ge surface, and that the uphill edge of GNRs is H-terminated and unpinned from the Ge surface. These first studies on the GQDs and GNRs growth on Ge surfaces provide feasible approaches to achieve unidirectional GNRs' growth on CMOS-compatible substrates, like $3\mu\mathrm{m}$-Ge(001)/Si(001), which may substantially promote GNRs scalable integration into the future semiconductor technology.

\section*{Conclusions}

Our minireview presents a timely and comprehensive look on the new topic in graphene and semiconductors research. Over the last several years, since the first publications on the successful graphene synthesis on Ge surfaces, many experimental works were carried out on the growth and electronic structure studies of different gr/Ge interfaces. These systematic works are mainly devoted to the studies of the growth mechanisms as well as on the finding of the experimental conditions for the growth of highly ordered and high quality graphene layers on Ge surfaces. It is found that Ge(110) surface is the most suitable one for the preparation of high-quality graphene layers and the most technologically attractive Ge(001) undergoes strong surface faceting during graphene preparation (which can be released via high temperature annealing either in UHV or under inert atmosphere). First electronic structure studies by means of NEXAFS and ARPES indicate the formation of free-standing graphene with moderate $n$-doping. Furthermore, it is shown, that alike the gr/metal interfaces, one can perform successful modification of the properties of gr/Ge interfaces via intercalation of different species. In cases of hydrogen and oxygen intercalation, this leads to the release of the initial compressive strain in graphene with formation of almost charge neutral graphene. One of the most interesting topics in gr/Ge studies is the preparation and electronic structure studies of different low-dimensional objects, like GQDs and GNRs, which growth and orientation on Ge is dictated by the substrate orientation (for example, flat or vicinal). 

Overall, the present status of the gr/Ge research and recent results can present a successful starting step toward a possible future integration of graphene in the modern semiconductors technology. Here, two possible strategies for this can be considered. In the first case, graphene is grown on Ge via metal-contamination-free synthesis and then transferred on the desired support using, i.\,e., poly methyl methacrylate (PMMA) polymer. Here, one will avoid the contamination with residual metal atoms, however organic contaminations coming from the transfer process and the residual contaminations at the gr-Ge interface still have to be removed via additional low-temperature annealing process. (See Ref.~\citenum{Backes:2020ed} for the recent review of the different processing methods for 2D materials.) In the second case, the direct processing of the metal-free synthesised gr/Ge systems is assumed. Here, the use of chemical reagents and polymers is avoided. However, the main factor, which define the quality of a graphene layer and gr/Ge interface, remains the high temperature used during graphene synthesis. As was shown, the temperature used during preparation of the high quality graphene layers on Ge is very close to the melting temperature that can lead to the unavoidable Ge sublimation and possible contamination of the synthesis facility, modifications of the implanted regions in the semiconductors substrates, degradation of the already fabricated structures, etc. Also high temperature used during synthesis in both cases, can lead to the dopants segregation at the gr/Ge interface that might lead to the uncontrollable modifications of the properties of the synthesised graphene. In case of the GNDs and GNRs growth the possibility of the space selective deposition is also highly desirable.

All described problems and concerns point out the significance of further experimental and theoretical studies of different important aspects, like, e.\,g., taking into account the substrate doping, controllable modifications of the gr/Ge interfaces via intercalation of different atoms, adsorption of different species on top (to mimic the environmental conditions), and creation of different gr/Ge-based heterostructures for the \textit{in operando} studies.

\section*{Conflicts of interest}

There are no conflicts to declare.

\section*{Acknowledgements}

This work was supported by the National Natural Science Foundation of China (Grant No. 21973059). Y.\,D. and E.\,V. thanks the financial support by the Ministry of Science and Higher Education of the Russian Federation (State assignment in the field of scientific activity, Southern Federal University, 2020).


\balance


\bibliographystyle{rsc} 

\begin{thebibliography}{99}

\bibitem{Wallace:1947dy}
P.~R. Wallace, \emph{Phys. Rev.}, 1947, \textbf{71}, 622--634\relax
\bibitem{Semenoff:1984dq}
G.~W. Semenoff, \emph{Phys. Rev. Lett.}, 1984, \textbf{53}, 2449--2452\relax
\bibitem{Schakel:1991aa}
A.~M.~J. Schakel and G.~W. Semenoff, \emph{Phys. Rev. Lett.}, 1991,
  \textbf{66}, 2653--2656\relax
\bibitem{Hohenberg:1967aa}
{Hohenberg, P. C.}, \emph{Phys. Rev.}, 1967, \textbf{158}, 383--386\relax
\bibitem{Mermin:1968aa}
N.~D. Mermin, \emph{Phys. Rev.}, 1968, \textbf{176}, 250--254\relax
\bibitem{Novoselov:2004a}
K.~Novoselov, A.~Geim, S.~Morozov, D.~Jiang, Y.~Zhang, S.~Dubonos,
  I.~Grigorieva and A.~Firsov, \emph{Science}, 2004, \textbf{306},
  666--669\relax
\bibitem{Novoselov:2005es}
K.~Novoselov, A.~Geim, S.~Morozov, D.~Jiang, M.~Katsnelson, I.~Grigorieva,
  S.~Dubonos and A.~Firsov, \emph{Nature}, 2005, \textbf{438}, 197--200\relax
\bibitem{Zhang:2005gp}
Y.~Zhang, Y.~Tan, H.~Stormer and P.~Kim, \emph{Nature}, 2005, \textbf{438},
  201--204\relax
\bibitem{Geim:2007hy}
A.~K. Geim and K.~S. Novoselov, \emph{Nature Mater.}, 2007, \textbf{6},
  183--191\relax
\bibitem{Geim:2009}
A.~Geim, \emph{Science}, 2009, \textbf{324}, 1530--1534\relax
\bibitem{Geim:2014hf}
A.~K. Geim and I.~V. Grigorieva, \emph{Nature}, 2014, \textbf{499},
  419--425\relax
\bibitem{Duan:2015fr}
X.~Duan, C.~Wang, A.~Pan, R.~Yu and X.~Duan, \emph{Chem. Soc. Rev.}, 2015,
  \textbf{44}, 8859--8876\relax
\bibitem{Manzeli:2017ib}
S.~Manzeli, D.~Ovchinnikov, D.~Pasquier, O.~V. Yazyev and A.~Kis,
  \emph{Nature}, 2017, \textbf{2}, 147--15\relax
\bibitem{Hagstrom:1965vh}
S.~Hagstrom, H.~B. Lyon and G.~A. Somorjai, \emph{Phys. Rev. Lett.}, 1965,
  \textbf{15}, 491\relax
\bibitem[May(1969)]{May:1969uj}
J.~W. May, \emph{Surf. Sci.}, 1969, \textbf{17}, 267--270\relax
\bibitem{Land:1992}
T.~A. Land, T.~Michely, R.~Behm, J.~C. Hemminger and G.~Comsa, \emph{Surf.
  Sci.}, 1992, \textbf{264}, 261--270\relax
\bibitem{Tontegode:1991ts}
A.~Y. Tontegode, \emph{Prog. Surf. Sci.}, 1991, \textbf{38}, 201--429\relax
\bibitem{Dedkov:2001}
Y.~S. Dedkov, A.~M. Shikin, V.~K. Adamchuk, S.~L. Molodtsov, C.~Laubschat,
  A.~Bauer and G.~Kaindl, \emph{Phys. Rev. B}, 2001, \textbf{64}, 035405\relax
\bibitem{Batzill:2012}
M.~Batzill, \emph{Surf. Sci. Rep.}, 2012, \textbf{67}, 83--115\relax
\bibitem{Dedkov:2015kp}
Y.~Dedkov and E.~Voloshina, \emph{J. Phys.: Condens. Matter}, 2015,
  \textbf{27}, 303002\relax
\bibitem{Karpan:2007}
V.~M. Karpan, G.~Giovannetti, P.~A. Khomyakov, M.~Talanana, A.~A. Starikov,
  M.~Zwierzycki, J.~v.~d. Brink, G.~Brocks and P.~J. Kelly, \emph{Phys. Rev.
  Lett.}, 2007, \textbf{99}, 176602\relax
\bibitem{Karpan:2008}
V.~M. Karpan, P.~A. Khomyakov, A.~A. Starikov, G.~Giovannetti, M.~Zwierzycki,
  M.~Talanana, G.~Brocks, J.~v.~d. Brink and P.~J. Kelly, \emph{Phys. Rev. B},
  2008, \textbf{78}, 195419\relax
\bibitem{Kubler:2005fm}
L.~Kubler, K.~A{\"\i}t-Mansour, M.~Diani, D.~Dentel, J.~L. Bischoff and
  M.~Derivaz, \emph{Phys. Rev. B}, 2005, \textbf{72}, 115319\relax
\bibitem{Dedkov:2008d}
Y.~S. Dedkov, M.~Fonin and C.~Laubschat, \emph{Appl. Phys. Lett.}, 2008,
  \textbf{92}, 052506\relax
\bibitem{Dedkov:2008e}
Y.~S. Dedkov, M.~Fonin, U.~Ruediger and C.~Laubschat, \emph{Appl. Phys. Lett.},
  2008, \textbf{93}, 022509\relax
\bibitem{Sutter:2010bx}
E.~Sutter, P.~Albrecht, F.~E. Camino and P.~Sutter, \emph{Carbon}, 2010,
  \textbf{48}, 4414--4420\relax
\bibitem{Weatherup:2015cx}
R.~S. Weatherup, L.~D'Arsie, A.~Cabrero-Vilatela, S.~Caneva, R.~Blume,
  J.~Robertson, R.~Schloegl and S.~Hofmann, \emph{J. Am. Chem. Soc.}, 2015,
  \textbf{137}, 14358--14366\relax
\bibitem{NDiaye:2009a}
A.~T. N'Diaye, T.~Gerber, C.~Busse, J.~Myslivecek, J.~Coraux and T.~Michely,
  \emph{New J. Phys.}, 2009, \textbf{11}, 103045\relax
\bibitem{Gerber:2013fa}
T.~Gerber, J.~Knudsen, P.~J. Feibelman, E.~Gr{\aa}n{\"a}s, P.~Stratmann,
  K.~Schulte, J.~N. Andersen and T.~Michely, \emph{ACS Nano}, 2013, \textbf{7},
  2020--2031\relax
\bibitem{Paschke:2019ei}
F.~Paschke, P.~Erler, V.~Enenkel, L.~Gragnaniello and M.~Fonin, \emph{ACS
  Nano}, 2019, \textbf{13}, 780--785\relax
\bibitem{Deng:2016ch}
D.~Deng, K.~S. Novoselov, Q.~Fu, N.~Zheng, Z.~Tian and X.~Bao, \emph{Nature
  Nanotechnol.}, 2016, \textbf{11}, 218--230\relax
\bibitem{Fu:2016gt}
Q.~Fu and X.~Bao, \emph{Chem. Soc. Rev.}, 2017, \textbf{46}, 1842--1874\relax
\bibitem{Li:2009}
X.~Li, W.~Cai, J.~An, S.~Kim, J.~Nah, D.~Yang, R.~Piner, A.~Velamakanni,
  I.~Jung, E.~Tutuc, S.~K. Banerjee, L.~Colombo and R.~S. Ruoff,
  \emph{Science}, 2009, \textbf{324}, 1312--1314\relax
\bibitem{Kim:2009a}
K.~S. Kim, Y.~Zhao, H.~Jang, S.~Y. Lee, J.~M. Kim, K.~S. Kim, J.-H. Ahn,
  P.~Kim, J.-Y. Choi and B.~H. Hong, \emph{Nature}, 2009, \textbf{457},
  706--710\relax
\bibitem{Bae:2010}
S.~Bae, H.~Kim, Y.~Lee, X.~Xu, J.-S. Park, Y.~Zheng, J.~Balakrishnan, T.~Lei,
  H.~R. Kim, Y.~I. Song, Y.-J. Kim, K.~S. Kim, B.~Ozyilmaz, J.-H. Ahn, B.~H.
  Hong and S.~Iijima, \emph{Nat. Nanotech.}, 2010, \textbf{5}, 574--578\relax
\bibitem{Ryu:2014fo}
J.~Ryu, Y.~Kim, D.~Won, N.~Kim, J.~S. Park, E.-K. Lee, D.~Cho, S.-P. Cho, S.~J.
  Kim, G.~H. Ryu, H.-A.-S. Shin, Z.~Lee, B.~H. Hong and S.~Cho, \emph{ACS
  Nano}, 2014, \textbf{8}, 950--956\relax
\bibitem{Lupina:2015je}
G.~Lupina, J.~Kitzmann, I.~Costina, M.~Lukosius, C.~Wenger, A.~Wolff,
  S.~Vaziri, M.~{\"O}stling, I.~Pasternak, A.~Krajewska, W.~Strupinski,
  S.~Kataria, A.~Gahoi, M.~C. Lemme, G.~Ruhl, G.~Zoth, O.~Luxenhofer and
  W.~Mehr, \emph{ACS Nano}, 2015, \textbf{9}, 4776--4785\relax
\bibitem{Ambrosi:2014gl}
A.~Ambrosi and M.~Pumera, \emph{Nanoscale}, 2014, \textbf{6}, 472--476\relax
\bibitem{XiaohongAn:2013kq}
X.~An, F.~Liu, Y.~J. Jung and S.~Kar, \emph{Nano Letters}, 2013, \textbf{13},
  909--916\relax
\bibitem{Liu:2014daa}
F.~Liu and S.~Kar, \emph{ACS Nano}, 2014, \textbf{8}, 10270--10279\relax
\bibitem{HyeYoungKim:2013bj}
H.-Y. Kim, K.~Lee, N.~McEvoy, C.~Yim and G.~S. Duesberg, \emph{Nano Letters},
  2013, \textbf{13}, 2182--2188\relax
\bibitem{Singh:2013dh}
A.~Singh, M.~A. Uddin, T.~Sudarshan and G.~Koley, \emph{Small}, 2013,
  \textbf{10}, 1555--1565\relax
\bibitem{Gu:2012jo}
T.~Gu, N.~Petrone, J.~F. McMillan, A.~van~der Zande, M.~Yu, G.~Q. Lo, D.~L.
  Kwong, J.~Hone and C.~W. Wong, \emph{Nature Photonics}, 2012, \textbf{6},
  554--559\relax
\bibitem{Liu:2011ex}
M.~Liu, X.~Yin, E.~Ulin-Avila, B.~Geng, T.~Zentgraf, L.~Ju, F.~Wang and
  X.~Zhang, \emph{Nature}, 2011, \textbf{474}, 64--67\relax
\bibitem{Li:2010iw}
X.~Li, H.~Zhu, K.~Wang, A.~Cao, J.~Wei, C.~Li, Y.~Jia, Z.~Li, X.~Li and D.~Wu,
  \emph{Adv. Mater.}, 2010, \textbf{22}, 2743--2748\relax
\bibitem{An:2013gj}
X.~An, F.~Liu and S.~Kar, \emph{Carbon}, 2013, \textbf{57}, 329--337\relax
\bibitem{DiBartolomeo:2016ii}
A.~Di~Bartolomeo, \emph{Physics Reports}, 2016, \textbf{606}, 1--58\relax
\bibitem{Hackley:2009bf}
J.~Hackley, D.~Ali, J.~DiPasquale, J.~D. Demaree and C.~J.~K. Richardson,
  \emph{Appl. Phys. Lett.}, 2009, \textbf{95}, 133114\relax
\bibitem{ThanhTrung:2013io}
P.~Thanh~Trung, F.~Joucken, J.~Campos-Delgado, J.-P. Raskin, B.~Hackens and
  R.~Sporken, \emph{Appl. Phys. Lett.}, 2013, \textbf{102}, 013118\relax
\bibitem{ThanhTrung:2014hd}
P.~Thanh~Trung, J.~Campos-Delgado, F.~Joucken, J.-F. Colomer, B.~Hackens, J.-P.
  Raskin, C.~N. Santos and S.~Robert, \emph{J. Appl. Phys.}, 2014,
  \textbf{115}, 223704\relax
\bibitem{Maeda:2011bt}
F.~Maeda and H.~Hibino, \emph{Jpn. J. Appl. Phys.}, 2011, \textbf{50},
  06GE12\relax
\bibitem{Kim:2011fs}
K.-B. Kim, C.-M. Lee and J.~Choi, \emph{J. Phys. Chem. C}, 2011, \textbf{115},
  14488--14493\relax
\bibitem{Tai:2018fx}
L.~Tai, D.~Zhu, X.~Liu, T.~Yang, L.~Wang, R.~Wang, S.~Jiang, Z.~Chen, Z.~Xu and
  X.~Li, \emph{Nano-Micro Lett.}, 2018, \textbf{10}, 20\relax
\bibitem{Wang:2013fq}
G.~Wang, M.~Zhang, Y.~Zhu, G.~Ding, D.~Jiang, Q.~Guo, S.~Liu, X.~Xie, P.~K.
  Chu, Z.~Di and X.~Wang, \emph{Sci. Rep.}, 2013, \textbf{3}, 2465\relax
\bibitem{Lee:2014dv}
J.-H. Lee, E.~K. Lee, W.-J. Joo, Y.~Jang, B.-S. Kim, J.~Y. Lim, S.-H. Choi,
  S.~J. Ahn, J.~R. Ahn, M.-H. Park, C.-W. Yang, B.~L. Choi, S.~W. Hwang and
  D.~Whang, \emph{Science}, 2014, \textbf{344}, 286--289\relax
\bibitem{Lippert:2014fc}
G.~Lippert, J.~Dabrowski, T.~Schroeder, M.~A. Schubert, Y.~Yamamoto,
  F.~Herziger, J.~Maultzsch, J.~Baringhaus, C.~Tegenkamp, M.~C. Asensio,
  J.~Avila and G.~Lupina, \emph{Carbon}, 2014, \textbf{75}, 104--112\relax
\bibitem{Kiraly:2015kaa}
B.~Kiraly, R.~M. Jacobberger, A.~J. Mannix, G.~P. Campbell, M.~J. Bedzyk, M.~S.
  Arnold, M.~C. Hersam and N.~P. Guisinger, \emph{Nano Lett.}, 2015,
  \textbf{15}, 7414--7420\relax
\bibitem{Dabrowski:2017gr}
P.~Dabrowski, M.~Rogala, I.~Pasternak, J.~Baranowski, W.~Strupinski,
  M.~Kopciuszynski, R.~Zdyb, M.~Jalochowski, I.~Lutsyk and Z.~Klusek,
  \emph{Nano Res.}, 2017, \textbf{3}, 11700\relax
\bibitem{Sitek:2020hw}
J.~Sitek, I.~Pasternak, J.~Grzonka, J.~Sobieski, J.~Judek, P.~Dabrowski,
  M.~Zdrojek and W.~Strupinski, \emph{Appl. Surf. Sci.}, 2020, \textbf{499},
  143913\relax
\bibitem{Rogge:2015hl}
P.~C. Rogge, M.~E. Foster, J.~M. Wofford, K.~F. McCarty, N.~C. Bartelt and
  O.~D. Dubon, \emph{MRS Commun.}, 2015, \textbf{5}, 539--546\relax
\bibitem{DiGaspare:2018ha}
L.~Di~Gaspare, A.~M. Scaparro, M.~Fanfoni, L.~Fazi, A.~Sgarlata,
  A.~Notargiacomo, V.~Miseikis, C.~Coletti and M.~De~Seta, \emph{Carbon}, 2018,
  \textbf{134}, 183--188\relax
\bibitem{Tesch:2018hm}
J.~Tesch, F.~Paschke, M.~Fonin, M.~Wietstruk, S.~B{\"o}ttcher, R.~J. Koch,
  A.~Bostwick, C.~Jozwiak, E.~Rotenberg, A.~Makarova, B.~Paulus, E.~Voloshina
  and Y.~Dedkov, \emph{Nanoscale}, 2018, \textbf{10}, 6088--6098\relax
 \bibitem{Persichetti:2019hn}
L.~Persichetti, L.~Di~Gaspare, F.~Fabbri, A.~M. Scaparro, A.~Notargiacomo,
  A.~Sgarlata, M.~Fanfoni, V.~Miseikis, C.~Coletti and M.~De~Seta,
  \emph{Carbon}, 2019, \textbf{145}, 345--351\relax
\bibitem{Persichetti:2020il}
L.~Persichetti, M.~De~Seta, A.~M. Scaparro, V.~Miseikis, A.~Notargiacomo,
  A.~Ruocco, A.~Sgarlata, M.~Fanfoni, F.~Fabbri, C.~Coletti and L.~Di~Gaspare,
  \emph{Appl. Surf. Sci.}, 2020, \textbf{499}, 143923\relax
\bibitem{McElhinny:2016gw}
K.~M. McElhinny, R.~M. Jacobberger, A.~J. Zaug, M.~S. Arnold and P.~G. Evans,
  \emph{Surf. Sci.}, 2016, \textbf{647}, 90--95\relax
\bibitem{Pasternak:2016ec}
I.~Pasternak, P.~Dabrowski, P.~Ciepielewski, V.~Kolkovsky, Z.~Klusek, J.~M.
  Baranowski and W.~Strupi{\'{n}}ski, \emph{Nanoscale}, 2016, \textbf{8},
  11241--11247\relax
\bibitem{Lukosius:2016ce}
M.~Lukosius, J.~Dabrowski, J.~Kitzmann, O.~Fursenko, F.~Akhtar, M.~Lisker,
  G.~Lippert, S.~Schulze, Y.~Yamamoto, M.~A. Schubert, H.~M. Krause, A.~Wolff,
  A.~Mai, T.~Schroeder and G.~Lupina, \emph{ACS Appl. Mater. Interfaces}, 2016,
  \textbf{8}, 33786--33793\relax
\bibitem{Dai:2016jm}
J.~Dai, D.~Wang, M.~Zhang, T.~Niu, A.~Li, M.~Ye, S.~Qiao, G.~Ding, X.~Xie,
  Y.~Wang, P.~K. Chu, Q.~Yuan, Z.~Di, X.~Wang, F.~Ding and B.~I. Yakobson,
  \emph{Nano Letters}, 2016, \textbf{16}, 3160--3165\relax
\bibitem{Kiraly:2018db}
B.~Kiraly, A.~J. Mannix, R.~M. Jacobberger, B.~L. Fisher, M.~S. Arnold, M.~C.
  Hersam and N.~P. Guisinger, \emph{Appl. Phys. Lett.}, 2018, \textbf{113},
  213103\relax
\bibitem{Campbell:2018iia}
G.~P. Campbell, B.~Kiraly, R.~M. Jacobberger, A.~J. Mannix, M.~S. Arnold, M.~C.
  Hersam, N.~P. Guisinger and M.~J. Bedzyk, \emph{Phys. Rev. Materials}, 2018,
  \textbf{2}, 044004\relax
\bibitem{Zhou:2018ji}
D.~Zhou, Z.~Niu and T.~Niu, \emph{J. Phys. Chem. C}, 2018, \textbf{122},
  21874--21882\relax
\bibitem{Tesch:2017gm}
J.~Tesch, E.~Voloshina, M.~Fonin and Y.~Dedkov, \emph{Carbon}, 2017,
  \textbf{122}, 428--433\relax
\bibitem{Zhao:2019eh}
Y.~Zhao, D.~Han, X.~Wang, Z.~Hu, Y.~Chen, Y.~Chen, D.~Zhou, Y.~Li, E.~G. Fu and
  Z.~Zhao, \emph{Carbon}, 2019, \textbf{153}, 776--782\relax
\bibitem{Lee:2012gy}
J.~E. Lee, G.~Ahn, J.~Shim, Y.~S. Lee and S.~Ryu, \emph{Nat. Commun.}, 2012,
  \textbf{3}, 1024\relax
\bibitem{Dabrowski:2016im}
J.~Dabrowski, G.~Lippert, J.~Avila, J.~Baringhaus, I.~Colambo, Y.~S. Dedkov,
  F.~Herziger, G.~Lupina, J.~Maultzsch, T.~Schaffus, T.~Schroeder, M.~Kot,
  C.~Tegenkamp, D.~Vignaud and M.~C. Asensio, \emph{Sci. Rep.}, 2016,
  \textbf{6}, 31639\relax
\bibitem{Scaparro:2016jc}
A.~M. Scaparro, V.~Miseikis, C.~Coletti, A.~Notargiacomo, M.~Pea, M.~De~Seta
  and L.~Di~Gaspare, \emph{ACS Appl. Mater. Interfaces}, 2016, \textbf{8},
  33083--33090\relax
\bibitem{Preobrajenski:2008}
A.~B. Preobrajenski, M.~L. Ng, A.~S. Vinogradov and N.~Martensson, \emph{Phys.
  Rev. B}, 2008, \textbf{78}, 073401\relax
\bibitem{Schroder:2016eb}
U.~A. Schr{\"o}der, M.~Petrovi{\'c}, T.~Gerber, A.~J. Martinez-Galera,
  E.~Gr{\aa}n{\"a}s, M.~A. Arman, C.~Herbig, J.~Schnadt, M.~Kralj, J.~Knudsen
  and T.~Michely, \emph{2D Materials}, 2017, \textbf{4}, 015013\relax
\bibitem{Voloshina:2013cw}
E.~Voloshina, R.~Ovcharenko, A.~Shulakov and Y.~Dedkov, \emph{J. Chem. Phys.},
  2013, \textbf{138}, 154706\relax
\bibitem{Weser:2010}
M.~Weser, Y.~Rehder, K.~Horn, M.~Sicot, M.~Fonin, A.~B. Preobrajenski, E.~N.
  Voloshina, E.~Goering and Y.~S. Dedkov, \emph{Appl. Phys. Lett.}, 2010,
  \textbf{96}, 012504\relax
\bibitem{Dedkov:2010jh}
Y.~S. Dedkov and M.~Fonin, \emph{New J. Phys.}, 2010, \textbf{12}, 125004\relax
\bibitem{Rusz:2010}
J.~Rusz, A.~B. Preobrajenski, M.~L. Ng, N.~A. Vinogradov, N.~Martensson,
  O.~Wessely, B.~Sanyal and O.~Eriksson, \emph{Phys. Rev. B}, 2010,
  \textbf{81}, 073402\relax
\bibitem{Ahn:2018fja}
S.~J. Ahn, H.~W. Kim, I.~B. Khadka, K.~B. Rai, J.~R. Ahn, J.-H. Lee, S.~G. Kang
  and D.~Whang, \emph{J. Korean Phys. Soc.}, 2018, \textbf{73}, 656--660\relax
\bibitem{Hwang:2012he}
C.~Hwang, D.~A. Siegel, S.-K. Mo, W.~Regan, A.~Ismach, Y.~Zhang, A.~Zettl and
  A.~Lanzara, \emph{Sci. Rep.}, 2012, \textbf{2}, 590\relax
\bibitem{Ryu:2017dx}
H.~Ryu, J.~Hwang, D.~Wang, A.~S. Disa, J.~Denlinger, Y.~Zhang, S.-K. Mo,
  C.~Hwang and A.~Lanzara, \emph{Nano Lett.}, 2017, \textbf{17},
  5914--5918\relax
\bibitem{Jacobberger:2019gk}
R.~M. Jacobberger, E.~A. Murray, M.~Fortin-Deschenes, F.~G{\"o}ltl, W.~A. Behn,
  Z.~J. Krebs, P.~L. L{\'e}vesque, D.~E. Savage, C.~Smoot, M.~G. Lagally,
  P.~Desjardins, R.~Martel, V.~Brar, O.~Moutanabbir, M.~Mavrikakis and M.~S.
  Arnold, \emph{Nanoscale}, 2019, \textbf{11}, 4864--4875\relax
\bibitem{Riedl:2009el}
C.~Riedl, C.~Coletti, T.~Iwasaki, A.~A. Zakharov and U.~Starke, \emph{Phys.
  Rev. Lett.}, 2009, \textbf{103}, 246804\relax
\bibitem{Riedl:2010du}
C.~Riedl, C.~Coletti and U.~Starke, \emph{J. Phys. D: Appl. Phys.}, 2010,
  \textbf{43}, 374009\relax
\bibitem{Sforzini:2015cp}
J.~Sforzini, L.~Nemec, T.~Denig, B.~Stadtm{\"u}ller, T.~L. Lee, C.~Kumpf,
  S.~Soubatch, U.~Starke, P.~Rinke, V.~Blum, F.~C. Bocquet and F.~S. Tautz,
  \emph{Phys. Rev. Lett.}, 2015, \textbf{114}, 106804\relax
\bibitem{Grzonka:2018aaa}
J.~Grzonka, I.~Pasternak, P.~P. Micha{\l}owski, V.~Kolkovsky and W.~Strupinski,
  \emph{Appl. Surf. Sci.}, 2018, \textbf{447}, 582--586\relax
\bibitem{Judek:2019ha}
J.~Judek, I.~Pasternak, P.~Dabrowski, W.~Strupinski and M.~Zdrojek, \emph{Appl.
  Surf. Sci.}, 2019, \textbf{473}, 203--208\relax
\bibitem{BraeuningerWeimer:2019kx}
P.~Braeuninger-Weimer, O.~Burton, R.~S. Weatherup, R.~Wang, P.~Dudin,
  B.~Brennan, A.~J. Pollard, B.~C. Bayer, V.~P. Veigang-Radulescu, J.~C. Meyer,
  B.~J. Murdoch, P.~J. Cumpson and S.~Hofmann, \emph{APL Mater.}, 2019,
  \textbf{7}, 071107\relax
\bibitem{Dabrowski:2019hj}
P.~Dabrowski, M.~Rogala, I.~Pasternak, P.~Krukowski, J.~M. Baranowski,
  W.~Strupi{\'{n}}ski, I.~Lutsyk, D.~A. Kowalczyk, S.~Paw{\l}owski and
  Z.~Klusek, \emph{Carbon}, 2019, \textbf{149}, 290--296\relax
\bibitem{Mendoza:2019de}
C.~D. Mendoza, M.~E.~H. Maia~da Costa and F.~L. Freire~Jr, \emph{Appl. Surf.
  Sci.}, 2019, \textbf{497}, 143779\relax
\bibitem{Jacobberger:2015de}
R.~M. Jacobberger, B.~Kiraly, M.~Fortin-Deschenes, P.~L. L{\'e}vesque, K.~M.
  McElhinny, G.~J. Brady, R.~R. Delgado, S.~S. Roy, A.~Mannix, M.~G. Lagally,
  P.~G. Evans, P.~Desjardins, R.~Martel, M.~C. Hersam, N.~P. Guisinger and
  M.~S. Arnold, \emph{Nat. Commun.}, 2015, \textbf{6}, 8006\relax
\bibitem{Saraswat:2019ei}
V.~Saraswat, Y.~Yamamoto, H.-J. Kim, R.~M. Jacobberger, K.~R. Jinkins, A.~J.
  Way, N.~P. Guisinger and M.~S. Arnold, \emph{J. Phys. Chem. C}, 2019,
  \textbf{123}, 18445--18454\relax
\bibitem{Li:2019bo}
P.~Li, T.~Wang, Y.~Yang, Y.~Wang, M.~Zhang, Z.~Xue and Z.~Di, \emph{Phys.
  Status Solidi RRL}, 2019, \textbf{97}, 1900398\relax
\bibitem{Backes:2020ed}
C.~Backes, A.~M. Abdelkader, C.~Alonso, A.~Andrieux-Ledier, R.~Arenal,
  J.~Azpeitia, N.~Balakrishnan, L.~Banszerus, J.~Barjon, R.~Bartali,
  S.~Bellani, C.~Berger, R.~Berger, M.~M.~B. Ortega, C.~Bernard, P.~H. Beton,
  A.~Beyer, A.~Bianco, P.~B{\o}ggild, F.~Bonaccorso, G.~B. Barin, C.~Botas,
  R.~A. Bueno, D.~Carriazo, A.~Castellanos-Gomez, M.~Christian, A.~Ciesielski,
  T.~Ciuk, M.~T. Cole, J.~Coleman, C.~Coletti, L.~Crema, H.~Cun, D.~Dasler,
  D.~De~Fazio, N.~D{\'\i}ez, S.~Drieschner, G.~S. Duesberg, R.~Fasel, X.~Feng,
  A.~Fina, S.~Forti, C.~Galiotis, G.~Garberoglio, J.~M. Garcia, J.~A. Garrido,
  M.~Gibertini, A.~G{\"o}lzh{\"a}user, J.~G{\'o}mez, T.~Greber, F.~Hauke,
  A.~Hemmi, I.~Hern{\'a}ndez-Rodr{\'\i}guez, A.~Hirsch, S.~A. Hodge, Y.~Huttel,
  P.~U. Jepsen, I.~Jimenez, U.~Kaiser, T.~Kaplas, H.~Kim, A.~Kis, K.~Papagelis,
  K.~Kostarelos, A.~Krajewska, K.~Lee, C.~Li, H.~Lipsanen, A.~Liscio, M.~R.
  Lohe, A.~Loiseau, L.~Lombardi, M.~Francisca~L{\'o}pez, O.~Martin,
  C.~Mart{\'\i}n, L.~Mart{\'\i}nez, J.~{\'A}. Mart{\'\i}n-Gago,
  J.~Ignacio~Mart{\'\i}nez, N.~Marzari, {\'A}.~Mayoral, J.~McManus, M.~Melucci,
  J.~M{\'e}ndez, C.~Merino, P.~Merino, A.~P. Meyer, E.~Miniussi, V.~Miseikis,
  N.~Mishra, V.~Morandi, C.~Munuera, R.~Mu{\~n}oz, H.~Nolan, L.~Ortolani, A.~K.
  Ott, I.~Palacio, V.~Palermo, J.~Parthenios, I.~Pasternak, A.~Patane,
  M.~Prato, H.~Prevost, V.~Prudkovskiy, N.~Pugno, T.~Rojo, A.~Rossi,
  P.~Ruffieux, P.~Samor{\`\i}, L.~Schu{\'e}, E.~Setijadi, T.~Seyller,
  G.~Speranza, C.~Stampfer, I.~Stenger, W.~Strupinski, Y.~Svirko, S.~Taioli,
  K.~B.~K. Teo, M.~Testi, F.~Tomarchio, M.~Tortello, E.~Treossi, A.~Turchanin,
  E.~Vazquez, E.~Villaro, P.~R. Whelan, Z.~Xia, R.~Yakimova, S.~Yang, G.~R.
  Yazdi, C.~Yim, D.~Yoon, X.~Zhang, X.~Zhuang, L.~Colombo, A.~C. Ferrari and
  M.~Garcia-Hernandez, \emph{2D Materials}, 2020, \textbf{7}, 022001\relax
\end{thebibliography}

\end{document}